\documentclass{aastex}
\usepackage{emulateapj5}
\shorttitle{Spectra of Comet C/2002\,C1 (Ikeya-Zhang)}
\begin{document}
\title{A Search for N$_2^+$ in Spectra of Comet C/2002\,C1 (Ikeya-Zhang)}
\author{A. L. Cochran }
\affil{The University of Texas}
\affil{Astronomy Department, The University of Texas, Austin, TX 78712}
\email{anita@barolo.as.utexas.edu}
\begin{abstract}
We report low- and high-resolution spectra of comet C/2002\,C1
(Ikeya-Zhang) from McDonald Observatory.  The comet had a well-developed
ion tail including CO$^+$, CO$^+_2$, CH$^+$, and H$_{2}$O$^+$.
We used our high-resolution spectra to search for N$_2^+$.
None was detected and we placed upper limits
on N$_2^+$/CO$^+$ of $5.4\times10^{-4}$.  N$_2^+$
was detected in the low-resolution spectra but we show that this
emission was probably telluric in origin (if cometary, we derive 
N$_2^+$/CO$^+ = 5.5\times10^{-3}$, still very low). 
We discuss the implications for the
conditions in the early solar nebula of the non-detection of
N$_2^+$.  These depend on whether the H$_{2}$O ice was deposited
in the amorphous or crystalline form.  If H$_{2}$O was deposited
in its crystalline form, the detection of CO$^+$ but not N$_2^+$
has implications for H$_{2}$O/H$_2$ in the early solar nebula.
\end{abstract}

\section{INTRODUCTION}
Knowledge of the nitrogen content of comets is important for an
understanding of conditions in the early solar nebula. 
Nitrogen exists in cometary ices in many forms including N$_2$, NH$_{3}$, HCN,
etc. Most of these species are chemically reactive in the coma gases
making it difficult to use coma observations
to unravel the nitrogen chemistry of the solar nebula.
However, conditions in the early solar nebula
were such that  the dominant equilibrium species of carbon, oxygen and
nitrogen should be N$_2$, CO, and H$_{2}$O \citep{lepr80}.
Observations of N$_2$H$^+$ in dense molecular clouds, coupled with
chemical models, led \citet{wowyzi92} to infer that in potential
star-forming regions
nitrogen is preferentially in N$_2$, rather than NH$_{3}$.

N$_2$ is the least reactive of the nitrogen-bearing species and is thus
the most appropriate to study to understand the nitrogen chemistry
of comets.  However, observations of cometary N$_2$ are exceedingly difficult.
Ground-based spectra are hampered by the telluric N$_2$ atmosphere;
spacecraft flyby mass spectrometer observations are compromised by the fact that
N$_2$ shares the mass 28 bin with CO, which is known to be quite common in
cometary comae.

A suitable method for measuring the N$_2$ content of cometary ices is
by studying its ion, N$_2^+$.  Generally, observations are obtained of the
N$_2^+$ 1N $B^2\Sigma_u^+-X^1\Sigma_g^+$ (0,0) band at 3914\AA.
Successful observations of this band require that the comet have a
well-developed ion tail and that the data be obtained with sufficiently
high spectral resolution to isolate the band from other cometary
emissions and from any telluric N$_2^+$ emission.

Comet C/2002\,C1 (Ikeya-Zhang) was discovered in early February 2002 and
reached perihelion on 2002 March 18 at a heliocentric distance of
0.507\,{\sc au}.  Our spectral observations showed that it had a
strong ion tail containing ions of CO$^+$, CO$_2^+$, CH$^+$, and
H$_{2}$O$^+$.  The comet was relatively bright, making it possible
to observe at high spectral-resolution.  Thus, we obtained
spectra of Ikeya-Zhang at both high- and low-resolution in order
to search for the signature of N$_2^+$ in the tail.
In this paper, we report on our non-detection of any N$_2^+$
attributable to the comet and discuss the implications of our derived
upper limits.

\section{OBSERVATIONS}
We obtained spectra of comet Ikeya-Zhang during
April 2002 using the 2.7-m Harlan J. Smith Telescope at McDonald Observatory.
Table~\ref{log} gives the parameters of the observations.
Two instruments were used: the Large Cassegrain Spectrograph (LCS)
is a long-slit CCD instrument with a resolving power of
R=$\lambda/\Delta\lambda$=550; the 2dcoud\'{e} is a cross-dispersed echelle
spectrograph with R=60,000.  The LCS observations covered the spectral
bandpass from 3000--5600\AA.  The slit was 2\,arcsec wide and
approximately 2.5\,arcmin long, with each pixel imaging 1.28\,arcsec on
the sky.
The 2dcoud\'{e} observations covered the spectral
region from 3700--5700\AA\ continuously, with coverage to 1.02$\mu$m
with increasing interorder gaps. The slit was $1.2\times8.2$\,arcsec.
With both instruments, observations
were obtained of the optocenter, as well as of the tail.
The LCS data reduction procedure has been described in detail by
\citet{cobarast92}; the 2dcoud\'{e} data reduction was described
in \citet{cocoba00}.

\begin{figure}
\plotone{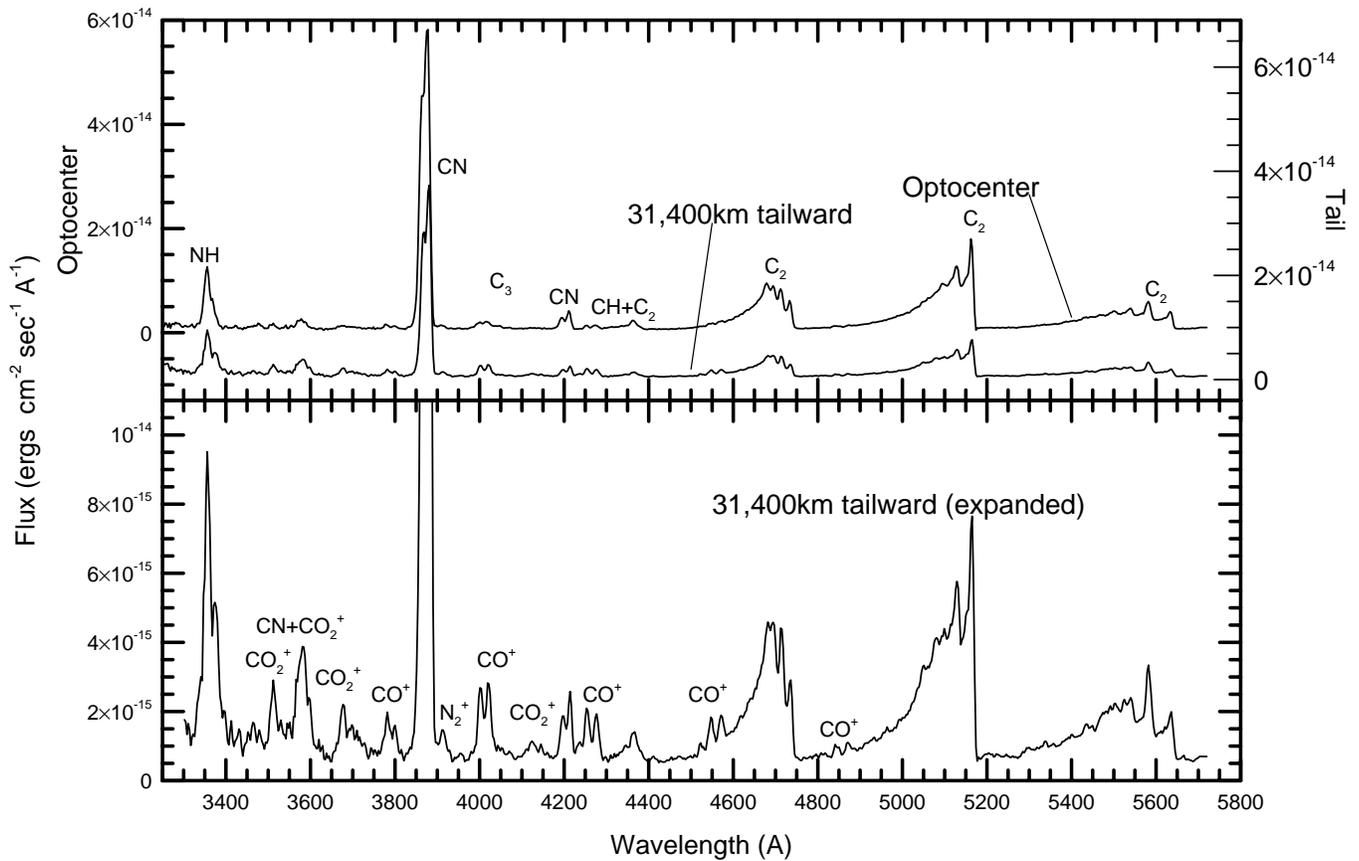}
\caption[fig1]{Low-resolution spectra of comet Ikeya-Zhang.  Shown
are spectra of the optocenter and 31,400\,km tailward of the optocenter.
In the upper panel, both spectra are shown to the same scale but
offset from one another (the scale for the optocenter is on the left;
that for the tail is on the right).  Many
features are marked.  In the lower panel the ordinate of the tail spectrum
has been expanded to show the weak ionic features.  The text discusses
the feature marked N$_2^+$.}\label{lowres}
\end{figure}

Figure~\ref{lowres} shows spectra obtained with the LCS,
both on the optocenter and 31,400\,km tailward.  The ion spectrum of
the tail is quite rich.  Note that there is a feature to the red of
the CN $\Delta$v=0 band which is labeled as N$_2^+$ in this figure. 
Is this really N$_2^+$ emission?
Some of it is certainly contamination from the CO$^+$ (5,1) band.
However, the relative fluorescence efficiencies in Table~3 of \citet{maah86}
can be used to predict that CO$^+$ from the (5,1) band contributes
no more than 1/4 of the detected feature.
Another possible contaminant is an unidentified
band of CO$_2^+$ with bandheads at 3906 and 3916\AA.
Evidence from \citet{mr47a,mr47b} suggests that the contamination
from this band is even less than that of the CO$^+$ (5,1) band.
Therefore, most of the observed feature at 3914\AA\ can be attributed
to N$_2^+$. 
Is this N$_2^+$ produced in the cometary coma or is it telluric in
origin?
We measured the integrated band fluxes of this feature, along with
the ionic CO$^+$ (3,0) and the neutral CN $\Delta$v=--1 bands
as a function of position within the coma.  While the CN and CO$^+$
features are fairly isolated, making them straightforward to measure,
the possible N$_2^+$ feature sits on the wing of the very strong
CN $\Delta$v=0 band.  Thus, extreme care must be taken when
removing the underlying ``continuum" of the feature.  This was done
by fitting the spectra at each position interactively.

\begin{figure}
\plotone{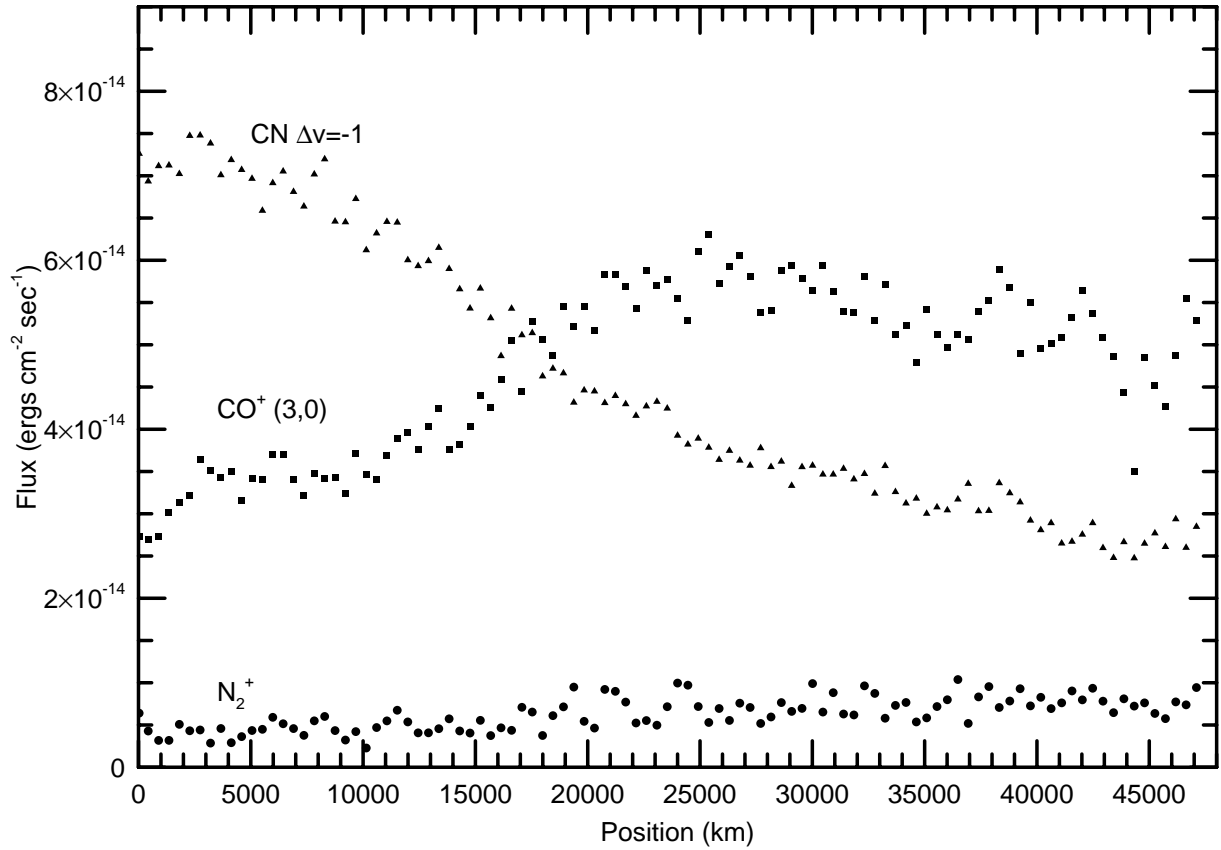}
\caption[fig2]{The distribution of the gas within the coma.  Shown
are examples of the distribution of a neutral (CN) and an ionized (CO$^+$)
species as a function of cometocentric distance.  In addition, the integrated
band flux of the feature denoted N$_2^+$ in Figure~\ref{lowres} is also
shown.  The distribution of this species is flat, unlike the CN or CO$^+$
gas.  The flat distribution is a signature of a telluric emission feature.}
\label{flux}
\end{figure}

Figure~\ref{flux} shows the derived integrated band fluxes
as a function of position within the coma.
The CN parent undergoes photodissociation as it flows outwards
from the nucleus.  In turn, the CN is photodissociated so that
there is a monotonic decrease
with cometocentric distance for the neutral CN. 
The CO$^+$ shows an increasing flux
until about 25,000\,km from the optocenter and then a slight decline.
This is a result of ionization within the coma and entrainment in the
solar wind.
The possible N$_2^+$ feature is mostly constant across the slit, unlike
either the neutral CN or the ionic CO$^+$. 
The S/N is low for this feature and removing the underlying CN adds
uncertainty.  If, as we believe, the feature is indeed constant across
the slit, it would be the result of the terrestrial atmosphere.
However, we cannot rule out that N$_2^+$ increases similarly to the CO+,
implying that there is a cometary contributor.

\begin{figure}
\plotone{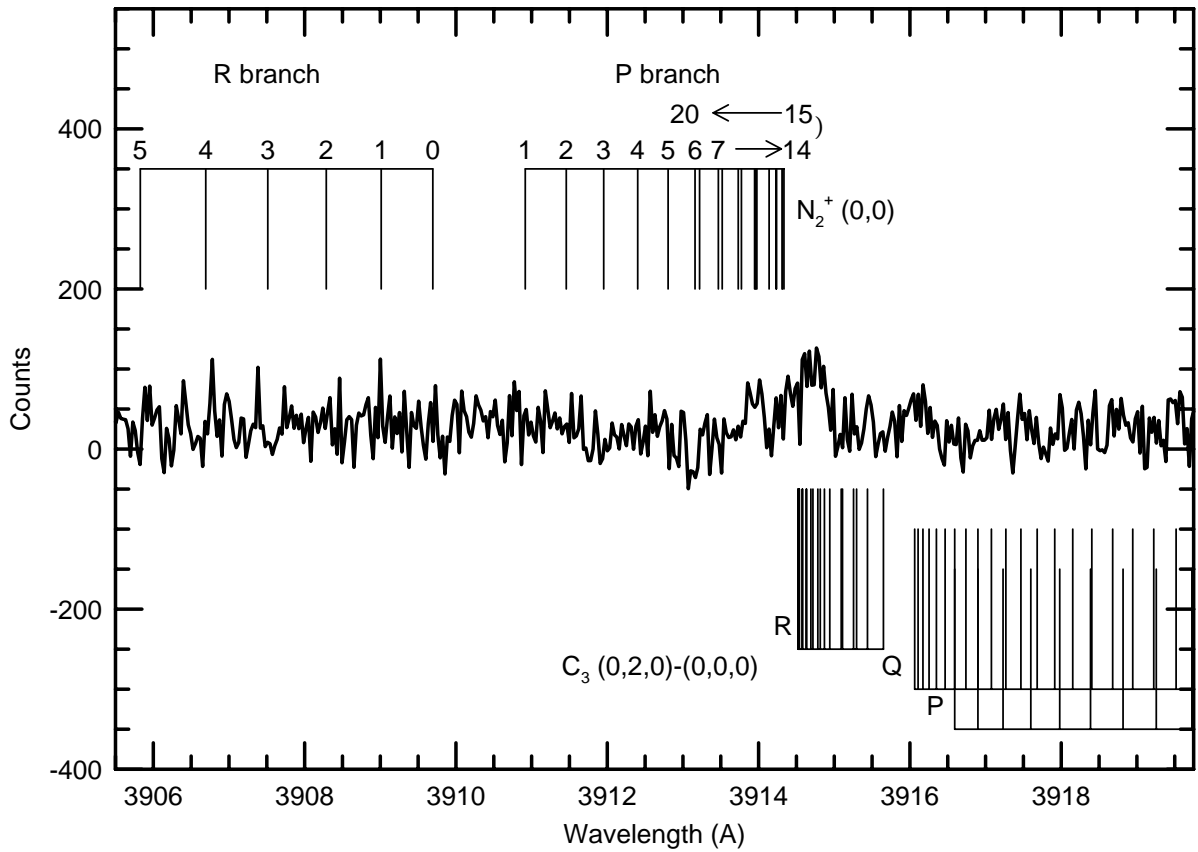}
\caption[fig3]{High-resolution spectra of comet Ikeya-Zhang.  Shown
is the spectrum of the tail in the region of the N$_2^+$ (0,0) band.
The expected positions of the P- and R-branch N$_2^+$ lines are marked.
None are detected.  Also marked are the expected positions of the
C$_{3}$ (0,2,0)--(0,0,0) band emissions.  The R- and Q-branches of
C$_{3}$ were detected, even 15,000\,km from the optocenter.}
\label{n2plus}
\end{figure}

High-resolution spectra should be more sensitive
than low-resolution spectra to weak line emission.
Therefore, on 2002 April 22, we obtained
echelle spectra to search for the N$_2^+$ (0,0) band.
Figure~\ref{n2plus} shows our tail spectrum of this spectral region.
The expected location of N$_2^+$ (0,0) band emission lines are marked
above the spectrum.  Inspection of the figure reveals no obvious
line emission at the wavelengths marked.  There is, however, an
obvious broad feature at around 3914.5\AA.  This feature begins just to 
the red of the N$_2^+$ P-branch bandhead.  A similar, but
weaker, feature was seen in comparable spectra of comet 122P/deVico
\citep{cocoba00}.  In that paper we argued that the
feature was the neutral C$_{3}$ (0,2,0)-(0,0,0) band.
In the deVico spectra, there was a single feature, while
close inspection of Figure~\ref{n2plus} reveals that in the
Ikeya-Zhang spectrum we have actually detected an additional feature at
around 3916\AA.  The first feature coincides with the C$_{3}$
R-branch bandhead while the second coincides with the Q-branch.
The presence of two of the C$_{3}$ (0,2,0)-(0,0,0) bandheads
confirms our identification of these features as C$_{3}$.
Indeed, comparison with the optocenter 2dcoud\'{e} spectra taken both
before and after the tail spectrum confirm this identification as well.

\section{UPPER LIMITS}
A limitation of the high-resolution 2dcoud\'{e} data is that it is not
easy to calibrate absolutely the spectra.  
However, we can calibrate the relative flux of different orders so
we can derive ratios of abundances of species for comparison
with models.  
To compute the upper limits of N$_2^+$/CO$^+$, we need to measure
an upper limit for the N$_2^+$ non-detection and the integrated
intensity of the CO$^+$ (2,0) band. 

For the integrated CO$^+$ intensity, we
fit a continuum to the region of the detected lines and summed
the counts in the band above the continuum.  Then, the contribution
of the strong CH$^+$ P(3) line was removed.  Finally, we
multiplied our integrated band intensity by two to reflect the
fact that we were observing only the $^2\Pi_{1/2}$(F$_2$) branches.
The integrated band intensity was 634 counts.

The N$_2^+$ upper limit is determined by asking how much signal
would we be able to hide within the noise.  We measured the
rms in the bandpass of the P-branch.  
Then, the
upper limit is just $1/2 \times \mathrm{rms} \times \mathrm{bandpass}$,
in appropriate units.
These are $2\sigma$ upper limits.
To account for the complete N$_2^+$ band, we multiplied the P-branch
upper limits by 2, assuming a comparable R-branch upper limit.
This results in a 2$\sigma$ upper limit of 4.2 counts.

To convert from the two band intensities to a ratio of the abundances,
we need to account for the relative response of the instrument
at the two bandpasses.  Using solar spectra obtained with the
same instrument, we determined that we needed to multiply
the counts in the order containing N$_2^+$ by a factor of 1.6.
In addition, we need to account for the relative fluorescence
efficiencies of the two species.
We used excitation factors of
$7.0\times10^{-2}$\,photons\,sec$^{-1}$\,mol$^{-1}$ for the N$_2^+$ (0,0) band
\citep{luwowa93} and $3.55\times10^{-3}$\,photons\,sec$^{-1}$\,mol$^{-1}$
for the CO$^{+}$ (2,0) band (the average value from Figure~2 of
\citet{maah86}).
Applying these factors, we find that
\[\frac{\mathrm N_2^+}{{\mathrm C\mathrm O}^+} = 5.4\times10^{-4}, \]
comparable to upper limits found for deVico and Hale-Bopp
with the same instrument \citep{cocoba00}.

Using the flux of the N$_2^+$ feature detected in the low-resolution
spectra, we can derive a ratio of
N$_2^+$/CO$^+ = 5.5\times10^{-3}$, a factor of 10 greater than our
upper limit derived from the high-resolution spectra.  This demonstrates
that we would have easily detected the signal seen in the low-resolution
spectra had it been present in the high-resolution spectra, regardless
of whether it was telluric or cometary.  If the N$_2^+$ in the low-resolution
spectra is cometary, it indicates a very variable parent.

If the N$_2^+$ in the low-resolution spectra was telluric, why didn't we
see it in the high-resolution spectra at the appropriate Doppler shift?
The N$_2^+$ $B^2\Sigma_u^+$ state requires 18.6eV to be populated.  Therefore,
in the Earth's atmosphere, it can only be found in dayglow or
auroral spectra.  There are no photochemical reactions which can produce
this emission in nightglow \citep{totori92}.  The geomagnetic $K_p$
indices for the nights of our observations were very similar
and showed relatively quiescent ionospheric conditions.  Thus, the presence
of the telluric N$_2^+$ in the low-resolution data 
cannot be the result of auroral activity.

The explanation for the presence of telluric N$_2^+$ in only some
of the spectra lies in the position of the comet in the sky at the
times of the observations.  While both the low- and high-resolution
spectra were obtained at approximately the same hour of the night,
around astronomical twilight, the orbit of the comet results in our
observing the comet at very different zenith distances during
the two observing runs.  During the early April timeframe of the
low-resolution data, the zenith distance of the comet was between 70 and
80 degrees; during the 2dcoud\'{e} observations, the zenith distance was
between 42 and 52 degrees.  Recall that the telluric N$_2^+$ emission
is a dayglow feature.  It is also excited high in the ionosphere,
with only 5\% of the N$_2^+$ vibrationally excited at 100\,km
and 50\% at 450\,km \citep{foda85}.  Thus,  we could only detect it
when the slant-path of the Sun illuminated the part of the atmosphere
we were observing through, or only at very high zenith distances.
Thus, if the N$_2^+$ was telluric, it is consistent that we only obseved
it in the low-resolution, high-zenith distance observations.

\section{IMPLICATIONS}
Ikeya-Zhang represents the third comet for which
we have obtained very high-quality, high-resolution spectra in which we
did not detect N$_2^+$ and could place tight constraints on
the relative abundances of N$_2^+$ and CO$^+$.  
Even if our low resolution N$_2^+$ detections were of cometary origin,we have
observed very little N$_2^+$ relative to CO$^+$ and this ratio is
variable for a single comet.
Previous detections and upper-limits from other observers were
discussed in \citet{cocoba00}.
In general, almost none of these previous observations were obtained 
at comparable resolutions and high sensitivity as our observations.

There is
an expectation that neutral N$_2$ and CO should be very common
in the early solar nebula \citep{lepr80} and thus, they are quite
likely deposited in ices in proportion to their source.
The quantity of these species deposited in the ices is dependent on several
factors, including the deposition temperature of the ice.
Current models of the dynamical evolution of the early Solar System
suggest that the Oort cloud comets we have studied were formed
in the Uranus-Neptune region prior to perturbation into the Oort cloud
(c.f. \citet{we91}; \citet{duqutr87}).
The temperature in this region was probably around $50\pm20$\,K
\citep{bomots89}.  Indeed, evidence confirms this deposition
temperature for Oort cloud comets \citep{meetal98b,meetal98a}.

The implications for the expected quantities of CO and N$_2$ ices in cometary
nuclei is then dependent on what form of H$_{2}$O ice is deposited
in the nucleus.  If the solar nebula ices never experienced warm
temperatures, the H$_{2}$O ice would be amorphous at the time of the
deposition of the nucleus.  Laboratory experiments of the deposition
of CO and N$_2$ gas in amorphous H$_{2}$O ices have shown that
CO is trapped 15--30 times more efficiently than N$_2$ in amorphous ice
if the temperature of the amorphous ice is around the temperatures
expected for the solar nebula \citep{baklko88,noba96}.
Using these results, \citet{owba95a} concluded that icy planetesimals
which formed in the Uranus-Neptune region would have N$_2$/CO$\approx 0.06$
in the gases trapped in the ice if N$_2$/CO$\approx 1$ in the nebula.
This value is much higher than any of our upper limits (or pssible
detection) for comets Ikeya-Zhang, deVico or Hale-Bopp.

\citet{moetal00} suggested that the H$_{2}$O was not deposited in its
amorphous state because of turbulence in the nebula at the time of deposition.  
Instead, the water ice infalling in the presolar cloud would first
have vaporized \citep{chca97} and subsequently been condensed in
the nebula in the form of crystalline ice \citep{koetal94}.
As the nebula continued to cool and H$_{2}$O ice crystallized, the other
volatiles would be trapped in hydrate clathrates \citep{iretal02}.

The trapping of the other volatiles via clathration is dependent
on the quantity of H$_{2}$O ice since clathration consumes water.
\citet{iretal02} pointed out that if H$_{2}$O/H$_2$ when water condensed
was less than 2.8 times the solar O/H ratio, the clathration process
would stop prior to trapping all of the volatiles.  Then, the ratio of
the trapped volatiles would be dependent on the curve of stability
for that species \citep{lust85}.  Inspection of Figure~2 of \citet{iretal02}
shows that CO would condense at warmer temperatures than N$_2$.
Thus, \citet{iretal02} suggested that our upper limits for N$_2^+$/CO$^+$
imply that H$_{2}$O/H$_2$ $<2.8\times$ solar O/H when deVico and
Hale-Bopp condensed.  Our new upper limit for N$_2^+$/CO$^+$
obtained for Ikeya-Zhang is consistent with this picture.

The three comets for which we have firm upper limits for N$_2^+$/CO$^+$
are consistent with formation of cometary nuclei from crystalline H$_{2}$O ice
and subsequent incomplete trapping of volatiles in hydrate clathrates.
All three of the comets are Oort cloud comets.  
\citet{iretal02} pointed out that conditions for comets
formed further from the proto-Sun, e.g., in the Kuiper belt, might
allow for the deposition of amorphous H$_{2}$O ice.  Thus,
it would be desirable to obtain high-resolution spectra of the ion tail
of any Kuiper belt comet which is bright enough in the future to test
our understanding of the conditions for deposition of ices in the outer
solar nebula.
However, such observations might not be conclusive because short-period
comets have undergone a great deal of volatile-depleting processing
during repeated solar passages.

\acknowledgements
This work was supported by NASA Grant NAG5-9003.  We thank Drs.
Michael Endl and Edwin Barker for their assistance obtaining some
of the spectra.


\begin{table}[ht]
\centering
\caption{Observational Circumstances}\label{log}
\begin{tabular}{r@{ }l@{ }rr@{--}lccccl}
 & & & \multicolumn{2}{c}{ } \\ [-9pt]
\hline
 & & & \multicolumn{2}{c}{ } \\ [-9pt]
\multicolumn{3}{c}{Date} & \multicolumn{2}{c}{UT Range} & R$_h$ &\.{R}$_h$ &
$\Delta$ & $\dot{\Delta}$ & Instrument \\
 & & & \multicolumn{2}{c}{ } & ({\sc {\sc au}}) & (km/sec) & ({\sc {\sc au}}) & (km/sec) \\
\hline
2002 & Apr & 9 & 10:44 & 11:30 & 0.70 & 26.4 & 0.51 & -16.4 & LCS \\
2002 & Apr & 10 & 10:44 & 11:29 & 0.72 & 26.8 & 0.50 & -15.6 & LCS \\
2002 & Apr & 11 & 11:03 & 11:30 & 0.74 & 27.2 & 0.49 & -14.9 & LCS \\
2002 & Apr & 22 & 10:20 & 11:40 & 0.92 & 29.2 & 0.42 & -6.5 & 2dcoud\'{e} \\
\hline
\end{tabular}
\end{table}

\end{document}